\documentclass[aps, prd, preprint,  11pt, superscriptaddress, nofootinbib]{revtex4-1}

\usepackage{fancyhdr}	
\usepackage{amsmath} \usepackage{amsfonts} \usepackage{amssymb}
\usepackage{wasysym}
\usepackage[dvipsnames]{xcolor}
\usepackage{graphicx}
\usepackage{multirow}
\usepackage{here}
\usepackage{epsfig}
\usepackage{epstopdf}
\usepackage{soul}
\usepackage{hyperref}
\usepackage{float}

\usepackage{array}
\newcolumntype{L}{>{\displaystyle}l}
\newcolumntype{C}{>{\displaystyle}c}
\newcolumntype{R}{>{\displaystyle}r}

\newcommand{\Rphi}{\mathcal{R}_{\delta\phi}}
\newcommand{\Rpsi}{\mathcal{R}_{\delta\psi}}

\newcommand{\Rm}{\mathcal{R}_{\delta\rho_\mathrm{m}}}

\renewcommand{\d}[1]{\mathinner{{\rm d}#1}}
\newcommand{\fn}[2]{\mathinner{#1\mathopen{\left(#2\right)}}}

\newcommand{\eq}[1]{Eq.~(\ref{#1})}
\newcommand{\eqs}[2]{Eqs.~(\ref{#1}) and (\ref{#2})}

\newcommand{\GeV}{\mathinner{\mathrm{GeV}}}

\newcommand{\invMpc}{\mathinner{\mathrm{Mpc}^{-1}}}

\newcommand{\fref}[1]{Figure~\ref{#1}}
\newcommand{\sref}[1]{Section~\ref{#1}}

\begin{document}

\title{CMB Spectral $\mu$-Distortion of Multiple Inflation Scenario}

\author{Gimin Bae}
\affiliation{School of Undergraduate Studies,
College of Transdisciplinary Studies,
Daegu Gyeongbuk Institute of Science and Technology (DGIST),
Daegu 42988, Republic of Korea}

\author{Sungjae Bae}
\affiliation{School of Undergraduate Studies,
College of Transdisciplinary Studies,
Daegu Gyeongbuk Institute of Science and Technology (DGIST),
Daegu 42988, Republic of Korea}

\author{Seungho Choe}
\email{schoe@dgist.ac.kr}
\affiliation{School of Undergraduate Studies,
College of Transdisciplinary Studies,
Daegu Gyeongbuk Institute of Science and Technology (DGIST),
Daegu 42988, Republic of Korea}

\author{Seo Hyun Lee}
\affiliation{School of Undergraduate Studies,
College of Transdisciplinary Studies,
Daegu Gyeongbuk Institute of Science and Technology (DGIST),
Daegu 42988, Republic of Korea}

\author{Jungwon Lim}
\affiliation{School of Undergraduate Studies,
College of Transdisciplinary Studies,
Daegu Gyeongbuk Institute of Science and Technology (DGIST),
Daegu 42988, Republic of Korea}

\author{Heeseung Zoe}
\email{heezoe@dgist.ac.kr}
\affiliation{School of Undergraduate Studies,
College of Transdisciplinary Studies,
Daegu Gyeongbuk Institute of Science and Technology (DGIST),
Daegu 42988, Republic of Korea}

\date{\today}

\begin{abstract}
In multiple inflation scenario having two inflations with an intermediate matter-dominated phase, the power spectrum is estimated to be enhanced on scales smaller than the horizon size at the beginning of the second inflation, $k > k_{\rm b}$.
We require $k_{\rm b} > 10 \invMpc$ to make sure that the enhanced power spectrum is consistent with large scale observation of cosmic microwave background (CMB). 
We consider the CMB spectral distortions generated by the dissipation of acoustic waves to constrain the power spectrum. 
The $\mu$-distortion value can be $10^2$ times larger than the expectation of the standard $\Lambda$CDM model ($\mu_{\Lambda\mathrm{CDM}} \simeq 2 \times 10^{-8}$) for $ k_{\rm b} \lesssim 10^3 \invMpc$, while the $y$-distortion is hardly affected by the enhancement of the power spectrum.
\end{abstract}

\maketitle


\section{Introduction}

Inflation provides the seeds of statistical fluctuations for the structure formation of the  universe \cite{gliner1, gliner2,Guth:1980zm,Linde:1981mu,Albrecht:1982wi} It fits with the large scale observations of the cosmic microwave background (CMB) and large scale structure (LSS) such as the
\emph{Wilkinson Microwave Anisotropy Probe} (WMAP) \cite{Spergel:2006hy}, the \emph{Planck} \cite{Ade:2015lrj} and \emph{Sloan Digital Sky Survey} (SDSS) \cite{Tegmark:2006az}.  However, there are many inflation models consistent with those large scale observations, and we should develop proper methods to specify the primordial inflation. One of the possible ways should be probing inflationary power spectrum on small scales by the observations such as ultracompact minihalos \cite{Bringmann:2011ut, Gosenca:2017ybi}, primordial black holes \cite{Carr:1975qj, Josan:2009qn}, the lensing dispersion of SNIa \cite{Ben-Dayan:2013eza, Ben-Dayan:2014iya, Ben-Dayan:2015zha}, the 21cm hydrogen line at or prior to the epoch of reionization \cite{Cooray:2006km, Mao:2008ug} or CMB distortions \cite{Chluba:2015bqa, Chluba:2011hw, Chluba:2012gq, Chluba:2012we}.

Multiple inflation scenario, having more than one inflationary periods after the first inflation, could leave characteristic signatures on small scales. Since the double inflation model, or inflation with a break, was introduced to give decoupling the power spectrum on large (CMB) and small (cluster-cluster/galaxy-galaxy) scales \cite{Silk:1986vc,Mukhanov:1991rp,Polarski:1992dq}, many versions of multiple inflation have been suggested as theoretical possibilities in supersymmetric particle physics models \cite{Adams:1997de,Kanazawa:1999ag,Lesgourgues:1999uc,Yamaguchi:2001pw,Burgess:2005sb,Kawasaki:2010ux,Lyth:1995hj, Lyth:1995ka, Hong:2015oqa, Cho:2017zkj,Hong:2017knn,Craig:2016lyx}.

CMB spectral distortions are a useful technique to probe small scales at $k \lesssim 10^4 \invMpc$ \cite{Chluba:2015bqa, Chluba:2011hw, Chluba:2012gq, Chluba:2012we}. COBE/FIRAS measurements indicate that the CMB photons are subject to the blackbody spectrum of temperature $T_\gamma = 2.726 \pm 0.001 {\rm K}$ with the spectral distortions $\Delta I / I \lesssim 5\times 10^{-5}$ \cite{Fixsen:1996nj,Mather:1993ij}. However, we have many astrophyscial/cosmological sources inducing spectral distortions: decaying or annihilating particles \cite{Chluba:2013wsa,Chluba:2013pya,Hu:1993gc,McDonald:2000bk,Sarkar:1984tt}, reionization and structure formation \cite{Cen:1998hc,Hu:1993tc,Miniati:2000iu,Oh:2003sa,Refregier:2000xz,SZ1972,Zhang:2003nr,Hill:2015tqa}, primordial black holes \cite{Carr:2009jm,Pani:2013hpa,Tashiro:2008sf}, cosmic strings \cite{Ostriker:1986xc,Tashiro:2012nb,Tashiro:2012nv,Tashiro:2012pp}, small-scale magnetic fields \cite{Jedamzik:1999bm,Kunze:2013uja,Sethi:2004pe,Chluba:2015lpa}, the adiabatic cooling of matter \cite{Chluba:2011hw, Ali-Haimoud:2015pwa}, cosmological recombination \cite{Dubrovich1975, Dubrovich1997, RubinoMartin:2006ug,Chluba:2006xa,SC2009,Chluba:2015gta}, gravitino decay \cite{Dimastrogiovanni:2015wvk}, and the dissipation of primordial density perturbations \cite{barrow,Chluba:2012we,Chluba:2013dna,Chluba:2012gq,Daly,Ganc:2012ae,Hu:1994bz,Pajer:2013oca,SZ1970,Clesse:2014pna,Emami:2015xqa,Dimastrogiovanni:2016aul,Chluba:2016aln} which is the main concern of this paper. Types of spectral distortions are characterized by the redshifts at which energy releases to CMB photons: At $z \gg 2 \times 10^6$, Compton and double Compton scatterings and Bremsstrahlung can effectively thermalize the energy release maintaining a black body spectrum \cite{SZmu, Hu:1993tc, Burigana}. At $ 3 \times 10^5 \lesssim z \lesssim 2 \times 10^6 $, only Compton scattering can efficiently redistribute the energy injected to the CMB. Compton scattering keeps the electrons and photons in kinetic equilibrium forming a chemical potential $\mu$. We define $\mu$-distortion as the spectral distortions associated with this chemical potential \cite{SZmu}. At $z \lesssim 10^4$, Compton scattering becomes inefficient, and the photons diffuse only a little in energy. We define $y$-distortion as the spectral distortion caused by the energy release at this epoch. It is also connected with the Sunyaev-Zeldovich effect on galaxy clusters \cite{SZ1969}.

\begin{figure}
\begin{center}
\includegraphics[width=0.85\textwidth]{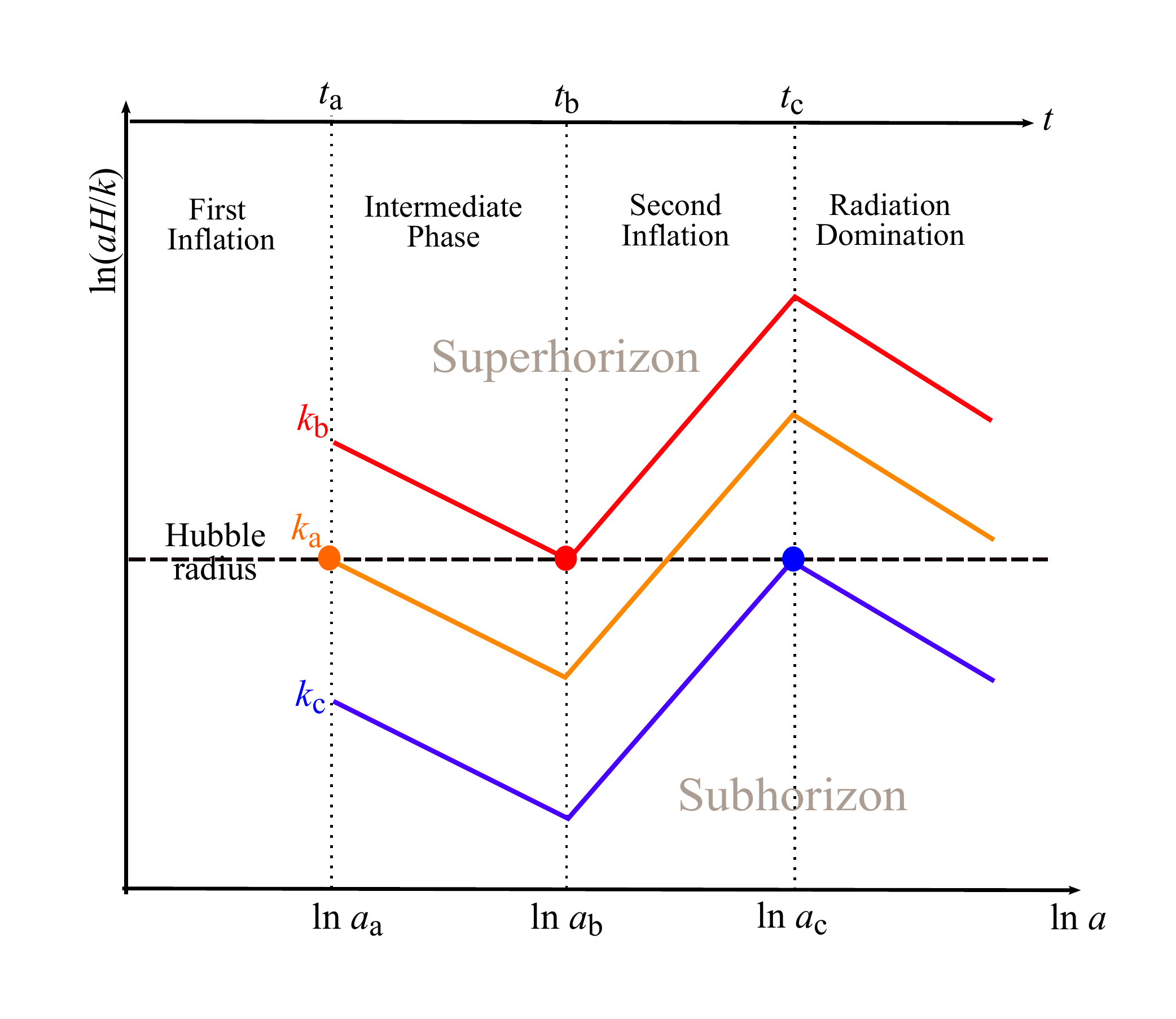}
\caption{Characteristic scales of multiple inflation with an intermediate phase between the primordial and the second inflations. Three characteristic scales $k_{\rm a}$, $k_{\rm b}$ and $k_{\rm c}$ correspond to the comoving scale of the horizon at each of the era boundaries. }
\label{scales}
\end{center}
\end{figure}

In this paper, we focus on \emph{multiple inflation with an intermediate matter-dominated period} (See \fref{scales}). We calculate the power spectrum, and predict the spectral distortion due to the dissipation of the density perturbations. In \cite{Cho:2017zkj,Nakama:2017ohe}, the power spectrum with suppression on small scales is discussed, and its implications on the spectral distortions are already mentioned. However, the power spectrum of multiple inflation with a matter-dominated break turns out to be enhanced on small scales, and its pattern of the spectral distortions should be different from them.

We discuss the evolution of the curvature perturbations generated by the multiple inflation scenario with an intermediate matter domination in \sref{PS}. In \sref{cmb_distortion}, we estimate the CMB spectral distortions by Silk damping of acoustic waves, due to the shear viscosity in the baryon-photon fluid. We find that the enhanced power spectrum of multiple inflation scenario could be constrained by the spectral distortion measurements. In \sref{dis}, we summarize our results and discuss possible ways to constrain  multiple inflation scenario.

\section{The evolution of cosmic perturbations}\label{PS}

\subsection{Overview}

We discuss the evolution of cosmological perturbations through two inflations with an intermediate phase as in \fref{scales}. The first inflation ends up with an intermediate phase, and then the second inflation begins after the intermediate phase and lasts until the usual radiation domination of Big-bang Nucleosynthesis (BBN) gets started.

There are many ways realizing these three stages. As an illustration, we consider a potential of two scalar fields $\fn{V}{\phi,\psi} = \fn{V_1}{\phi} + \fn{V_2}{\psi}$ having the following form.  We get the first inflation and the matter domination by $\fn{V_1}{\phi} = {V^*_1} - \frac{1}{2}m_{\phi}^2 \phi^2 + \cdots$. For $\phi \simeq 0$, the first inflation is driven by the vacuum energy $V_1^*$, but for $\phi > 0$, $\phi$ oscillates around its minimum with driving the matter phase. Note that  there is no reheating after the first inflation, if $\phi$ is not coupled with other fields. This is how we move from the first inflation to the matter domination without introducing an reheating phase between them. Then, we have the second inflation by $\fn{V_2}{\psi} = {V^*_2} + H^2 \psi^2 - \frac{1}{2}m_{\psi}^2 \psi^2 + \cdots$. For $H \simeq \left( V^*_1 \right)^{1/4} \gg m_{\psi}$, $\psi$ is safely held at the origin during the first inflation and the matter domination, and the second inflation starts as the matter energy density of $\phi$ is redshifted.

The characteristic scales are defined as
\begin{equation}
k_x \equiv a_x H_x
\end{equation}
where $a_x$ and $H_x$ are the scale factor and the Hubble parameter at the era boundary $t_x$. $k_x$ is the comoving scale of the horizon at $t=t_x$. The first inflation occurs at $t < t_{\rm a}$ generating the perturbations. While modes with  $k< k_{\rm b}$ remain outside the horizon before the radiation domination and are not affected by the phases after the first inflation, modes with $k > k_{\rm b}$ enter and exit the horizon and should evolve differently from those with $k \lesssim k_{\rm b}$ as the followings:
\begin{itemize}
\item Modes with $k_{\rm b} <  k < k_{\rm a} $ enter the horizon during the intermediate phase and then exit the horizon during the second inflation.

\item  Modes with $k_{\rm a} <  k < k_{\rm c} $ exit the horizon for the first time during the second inflation. 

\item Modes with $k > k_{\rm c} $ never exit the horizon and are not our interest. 
\end{itemize}
The power spectrum of the multiple inflation would fit with the large scale observation of CMB and LSS at $k< k_{\rm b}$, but should be clearly different from the expectations of a simple single primordial inflation or $\Lambda$CMD model at $k > k_{\rm b}$.   

Generic features of the power spectrum of multiple inflation with breaking intermediate periods are extensively studied in \cite{Namjoo:2012xs}. It is shown that the power spectrum is being enhanced with oscillations on small scales when the intermediate phase is matter dominated. Hence, we expect that the enhanced power spectrum could be explored by the CMB spctral distortions, and this is the motivation of this paper. In our paper, however, we adopt the scheme of multicomponent perturbation calculation in \cite{Hong:2015oqa} to refine the calculation on the curvature perturbation by making the background smoothly changing from the matter domination to the second inflation. From now on, we consider only matter domination as the intermediate phase, and discuss the power spectrum and its spectral distortions.

\subsection{During the first inflation}

The scalar part of the perturbed metric \cite{Kodama:1985bj} is 
\begin{equation}\label{metric}
\tilde{ds}^2 = (1+2A) \d{t}^2 - 2 B_{,i} \fn{a}{t}\d{t} \d{x^i} - \left[ (1+2\mathcal{R}) \fn{a^2}{t} \delta_{ij} + 2 C_{,ij} \right] \d{x^i} \d{x^j}
\end{equation}
We define 
\begin{equation}
\Rphi \equiv \mathcal{R} - \frac{H}{\dot{\phi}}\delta \phi
\end{equation}
where $\mathcal{R}$ is the intrinsic curvature perturbation on comoving hypersurfaces of \eq{metric} and $\phi$ is an inflaton field. Its mode functions are calculated by 
\begin{equation}\label{curpert}
\fn{\Rphi}{k,t} = - \frac{1}{2}\sqrt{\pi} e^{i\left(\nu + \frac{1}{2} \right)\frac{\pi}{2}} \left( \frac{1}{aH} \right)^\frac{1}{2} \left( \frac{H}{a\dot{\phi}} \right) \fn{H^{(1)}_\nu}{\frac{k}{aH}}~
\end{equation}
where $\nu = \frac{1+\delta + \epsilon}{1-\epsilon} + \frac{1}{2}$ with the slow-roll parameters  $\epsilon \equiv -\frac{\dot{H}}{H^2}$ and $\delta \equiv \frac{\ddot{\phi}}{H\dot{\phi}}$  (e.g. \citep{Stewart:1993bc}), and describe the evolution of the perturbation during the first inflation. From \eq{curpert}, we express the curvature perturbation at $t = t_{\rm a}$ as
\begin{equation}\label{Rphi}
\fn{{\Rphi}}{k,t_{\rm a}} = \frac{\sqrt{\pi}}{2} \fn{\alpha}{t_{\rm a}} H_{\rm a} \left( \frac{1}{k_{\rm a}} \right)^\frac{3}{2} \fn{H^{(1)}_\nu}{\frac{k}{k_{\rm a}}}
\end{equation}
where we assume $H \simeq H_{\rm a}$ throughout the first inflation and $\fn{\alpha}{t_{\rm a}} = \frac{H_{\rm a}}{\fn{\dot{\phi}}{t_{\rm a}}} $ is a constant depending on inflation models, $k_{\rm a} = a_{\rm a}H_{\rm a}$, and $\fn{H^{(1)}_\nu}{x}$ is the Hankel function of the first kind. 

For large scales, i.e. $k \ll k_{\rm a}$,
\begin{eqnarray}
\fn{{\Rphi}}{k,t_{\rm a}} \rightarrow 
-\frac{i \fn{ \Gamma}{\nu} 2^{\nu-1}}{\sqrt{\pi}} \fn{\alpha}{t_{\rm a}} \left( \frac{1}{k_{\rm a}} \right)^\frac{3}{2} \left( \frac{k_{\rm a}}{k} \right)^\nu
\end{eqnarray}
and the power spectrum is 
\begin{eqnarray}\label{Pphi}
\fn{P^{\frac{1}{2}}_{\Rphi}}{k} = 
\frac{2^{\nu-\frac{5}{2}}}{\pi} \frac{ \fn{ \Gamma}{\nu} }{\fn{ \Gamma}{\frac{3}{2}} } \fn{\alpha}{t_{\rm a}} \left( \frac{k}{k_{\rm a}} \right)^{\frac{3}{2}-\nu} 
\end{eqnarray}
where $\fn{\alpha}{t_{\rm a}}$ and $\nu$ will be fixed by the \emph{Planck} normalization \cite{Ade:2015lrj}.

\subsection{Matter-dominated intermediate phase}

We assume that the primordial universe can shift quickly from the first inflation to the matter domination. The evolution of the curvature perturbation on constant (matter) energy hypersurface during the matter-dominated intermediate phase is calculated by using the scheme of multicomponent perturbations introduced in \cite{Hong:2015oqa}\footnote{In \cite{Hong:2015oqa}, the authors discuss the evolution of curvature perturbation on constant energy hypersurface during the (moduli) matter domination. However, their result is very different from ours because they should calculate it in terms of radiation hypersurface which is relevant to describe the thermal inflation period.}. We assume that the second inflation vacuum energy becomes dominating as the matter energy density is gradually redshifted. The energy density is 
\begin{equation}\label{density}
\rho = \rho_{\rm m} + \rho_{\rm v}
\end{equation}
where $\rho_{\rm m}$ is the matter energy density and $\rho_{\rm v}$ is the vacuum energy density driving the second inflation. The characteristic scale $k_{\rm b}$ is naturally identified by the comoving scale at which the expansion rate of universe is changed
\begin{equation}\label{kb}
\fn{\ddot{a}}{t_{\rm b}} = 0~,
\end{equation}
and hence
\begin{equation}
k_{\rm b} \equiv a_{\rm b} H_{\rm b}~
\end{equation}
where $H_{\rm b}$ is the Hubble parameter during the second inflation and assumed to be constant. Note that \eqs{density}{kb} give 
\begin{equation}\label{kakb}
\frac{k_{\rm b}}{k_{\rm a}}= \frac{H_{\rm b}}{H_{\rm a}} \left( \frac{2}{3} \frac{H_{\rm a}^2}{H_{\rm b}^2}   -\frac{1}{2} \right)^{\frac{1}{3}}
\end{equation}
which produces the e-folds during the matter domination
\begin{equation}\label{aaab}
N_{\rm ab} = \log \frac{a_{\rm b}}{a_{\rm a}}= \log \left( \frac{2}{3} \frac{H_{\rm a}^2}{H_{\rm b}^2}   -\frac{1}{2} \right)^{\frac{1}{3}}~.
\end{equation}

The curvature perturbation on constant matter energy hypersurface is defined by
\begin{equation}
\Rm \equiv \mathcal{R} - \frac{H}{\dot{\rho}_{\rm m}}\delta \rho_{\rm m}~.
\end{equation}
whose mode functions should be matched with $\fn{\Rphi}{k,t}$ at $t = t_{\rm a}$ by requiring
\begin{eqnarray}
\fn{\Rm}{k,t_{\rm a}} & = & \fn{\Rphi}{k,t_{\rm a}} \\
\fn{\dot{\Rm}}{k,t_{\rm a}} & = & \fn{\dot{\Rphi}}{k,t_{\rm a}}
\end{eqnarray}

From \cite{Hong:2015oqa}, the governing equation is given by 
\begin{equation}\label{pert_m}
\ddot{\Rm} 
+ H \left( 2 + \frac{\rho_{\rm m} }{\rho_{\rm m} + \frac{2}{3}q^2} \right) \dot{\Rm}
- \left( \frac{q^2}{3} \right) \left( \frac{\rho_{\rm m} }{\rho_{\rm m} + \frac{2}{3}q^2} \right) \Rm = 0
\end{equation}
whose solution is reduced to
\begin{equation}
\fn{\Rm}{k,t_{\rm b}} = \fn{A_\mathrm{m}}{k,t_{\rm a}}  \left[ 1 + \fn{\mathcal{S}}{t_{\rm a}, t_{\rm b}} 
\left( \frac{k}{k_{\rm b}} \right)^2  \right]
+ \fn{B_\mathrm{m}}{k,t_\mathrm{a}} \left( \frac{H_{\rm b}}{H_{\rm a}} \right)
\end{equation}
where
\begin{equation}
\fn{\mathcal{S}}{t_{\rm a}, t_{\rm b}} = \frac{1}{3} 
\left( \frac{3}{2} \right)^\frac{3}{2} \int_{x_{\rm ab}}^{1} \left( \frac{ 2x }{2 + x^3} \right)^{\frac{3}{2}} d x
\end{equation}
with $x_{\rm ab} \equiv \frac{a_{\rm a}}{a_{\rm b}}$ and 
\begin{eqnarray}
\fn{A_\mathrm{m}}{k,t_{\rm a}} & = & \frac{1}{1+\frac{1}{3}\left( \frac{H_{\rm b}}{H_{\rm a}}  \right)\left( \frac{k_{\rm a}}{k_{\rm b}} \right) 
\left( \frac{k}{k_{\rm b}} \right)^2} 
\left[ \fn{\Rphi}{k,t_{\rm a}} 
+\left( \frac{H_{\rm b}}{H_{\rm a}} \right)
\left(\frac{k_{\rm a}}{k_{\rm b}}\right)^3 \frac{\fn{\dot\Rphi}{k,t_{\rm a}}}{{H_{\rm a}}} \right],
\end{eqnarray}
\begin{eqnarray}
\fn{B_\mathrm{m}}{k,t_{\rm a}} & = & \frac{\left( \frac{H_{\rm b}}{H_{\rm a}}  \right) \left( \frac{k_{\rm a}}{k_{\rm b}} \right) }{1+\frac{1}{3}\left( \frac{H_{\rm b}}{H_{\rm a}}  \right) \left( \frac{k_{\rm a}}{k_{\rm b}} \right) 
\left( \frac{k}{k_{\rm b}} \right)^2}  
\left[\frac{1}{3} 
\left( \frac{k}{k_{\rm b}} \right)^2 \fn{\Rphi}{k,t_{\rm a}} -
 \left( \frac{k_{\rm a}}{k_{\rm b}} \right)^2 \frac{\fn{\dot\Rphi}{k,t_{\rm a}}}{H_{\rm a}}
\right]~.
\end{eqnarray}

\subsection{The second inflation}

As the vacuum energy density of the second inflation gets dominating, the curvature perturbation 
\begin{equation}
\Rpsi \equiv \mathcal{R} - \frac{H}{\dot{\psi}}\delta \psi
\end{equation}
with $\psi$ the inflaton for the second inflation would be matched with $\Rm$ at $t= t_{\rm b}$ by  
\begin{eqnarray}
\fn{\Rpsi}{k,t_{\rm b}} & = & \fn{\Rm}{k,t_{\rm b}} \\
\fn{\dot{\Rpsi}}{k,t_{\rm b}} & = & \fn{\dot{\Rm}}{k,t_{\rm b}}
\end{eqnarray}

We treat the evolution of $\Rpsi$  in the same manner of the first inflation in \eq{curpert}, but have to consider both $H^{(1)}_\nu$ and $H^{(2)}_\nu$ for the second inflation as in \cite{Namjoo:2012xs}. We express the full solution by using Bessel functions, instead of Hankel functions, to have concise forms. 
\begin{equation}
\fn{\Rpsi}{k,t} = \left( \frac{k}{aH} \right)^{\nu'} \left[ \fn{C_{\psi}}{k,t_{\rm b}}  \fn{J_{\nu'}}{\frac{k}{aH}}
+ \fn{D_{\psi}}{k,t_{\rm b}}  \fn{Y_{\nu'}}{\frac{k}{aH}} \right].
\end{equation}
where $\nu' = \frac{1+\delta + \epsilon}{1-\epsilon} + \frac{1}{2}$ with the slow-roll parameters  $\epsilon \equiv -\frac{\dot{H}}{H^2}$ and $\delta \equiv \frac{\ddot{\psi}}{H\dot{\psi}}$ and
\begin{eqnarray}
\fn{C_{\psi}}{k,t_{\rm b}} & = & \frac{\pi}{2} \left( \frac{k_{\rm b}}{k}  \right)^{\nu'-1} 
\left[  
\fn{\Rm}{k,t_{\rm b}} \fn{Y_{\nu'-1}}{\frac{k}{k_{\rm b}}}  
+ \frac{\fn{\dot{\Rm}}{k,t_{\rm b}}}{H_{\rm b}} \left( \frac{k_{\rm b}}{k} \right)\fn{Y_{\nu'}}{\frac{k}{k_{\rm b}}}
\right] \\
\fn{D_{\psi}}{k,t_{\rm b}} & = & - \frac{\pi}{2} \left( \frac{k_{\rm b}}{k}  \right)^{\nu'-1} 
\left[  
\fn{\Rm}{k,t_{\rm b}} \fn{J_{\nu'-1}}{\frac{k}{k_{\rm b}}}  
+ \frac{\fn{\dot{\Rm}}{k,t_{\rm b}}}{H_{\rm b}} \left( \frac{k_{\rm b}}{k} \right)\fn{J_{\nu'}}{\frac{k}{k_{\rm b}}}
\right]\label{Dpsi}
\end{eqnarray}

Therefore, the curvature perturbation should be calculated at $t = t_{\rm c}$ by
\begin{equation}\label{Rpsi}
\fn{\Rpsi}{k,t_{\rm c}} = \left( \frac{k}{k_{\rm c}} \right)^{\nu'} \left[ \fn{C_{\psi}}{k,t_{\rm b}}  \fn{J_{\nu'}}{\frac{k}{k_{\rm c}}}
+ \fn{D_{\psi}}{k,t_{\rm b}}  \fn{Y_{\nu'}}{\frac{k}{k_{\rm c}}} \right].
\end{equation}

\subsection{Transfer Function and Power Spectrum}\label{TPS}

\begin{figure}
\begin{center}
\includegraphics[width=0.85\textwidth]{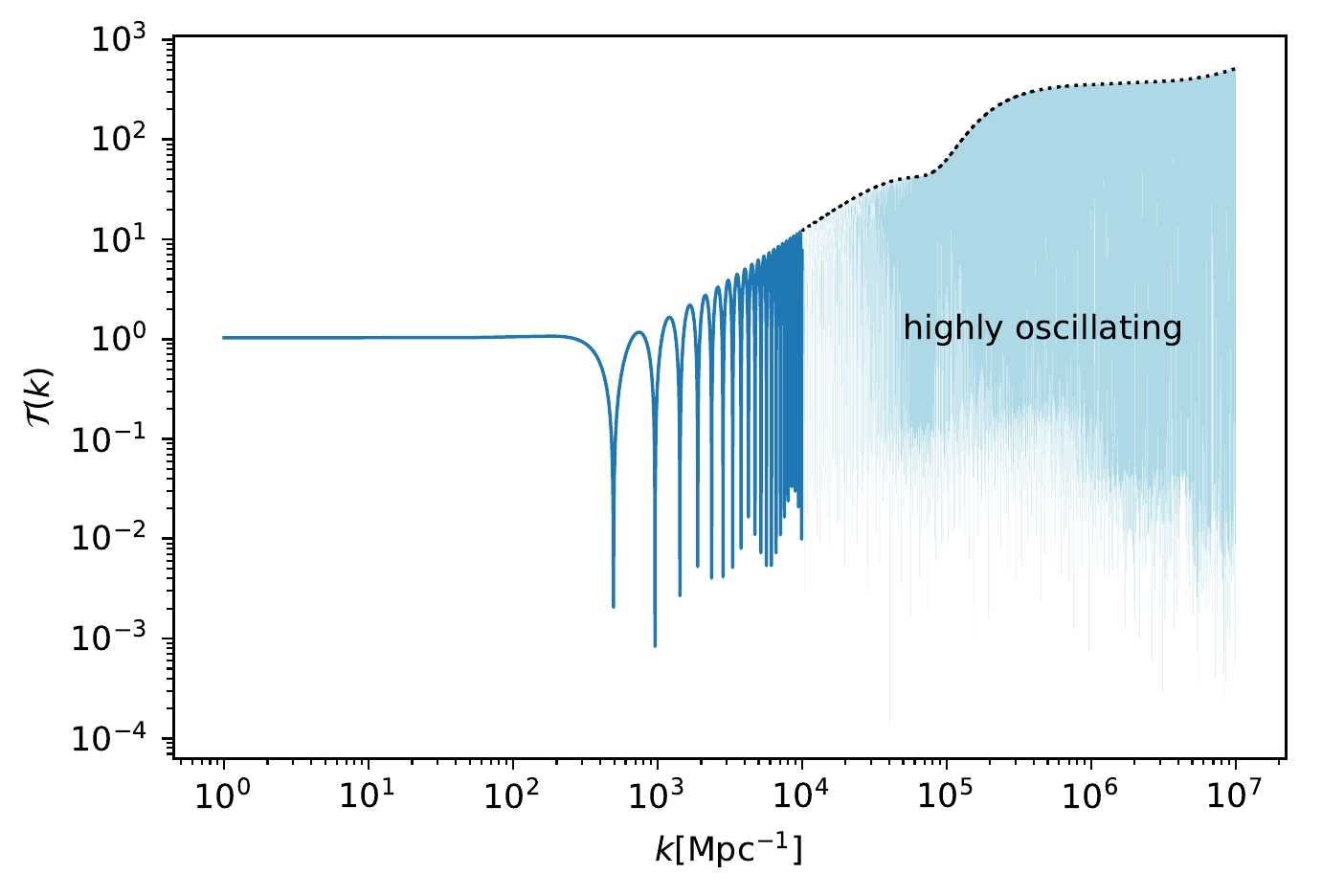}
\caption{Transfer Function $\fn{\mathcal{T}}{k}$ for $k_{\rm b} = 150 \invMpc$, $k_{\rm a} = 10^5 \invMpc$ and $k_{\rm c} = 10^7 \invMpc$.}
\label{transfer}
\end{center}
\end{figure}

The evolution of the curvature perturbation after the first inflation is summarized by a transfer function
\begin{equation}\label{transferfn}
\fn{\mathcal{T}}{k} \equiv \left| \frac{\fn{\Rpsi}{k,t_{\rm c}}}{\fn{\Rphi}{k,t_{\rm a}}} \right|
\end{equation}
and the resultant power spectrum is expressed by
\begin{equation}\label{Ppsi}
\fn{\mathcal{P}^{\frac{1}{2}}}{k} = \sqrt{\frac{k^3}{2\pi^2}} \left| \fn{\Rpsi}{k,t_{\rm c}} \right| = \sqrt{\frac{k^3}{2\pi^2}} \left| \fn{\mathcal{T}}{k}  \fn{\Rphi}{k,t_{\rm a}} \right|~.
\end{equation} 
For $k \ll k_{\rm b}$, the transfer function goes to $\mathcal{T} \rightarrow 1$, and the power spectrum is determined mainly by the first inflation,
We should take $\nu \simeq 1.52$ in \eq{Pphi} to follow the \emph{Planck} normalization $A^* = 2.21 \times 10^{-9} $ and $n_* = 0.96$ at $k_*= 0.05 \invMpc$ \citep{Ade:2015lrj}, and $\nu' \simeq 1.5$ in \eq{Rpsi} to include generic models for the second inflation.

\begin{figure}
\begin{center}
\includegraphics[width=0.85\textwidth]{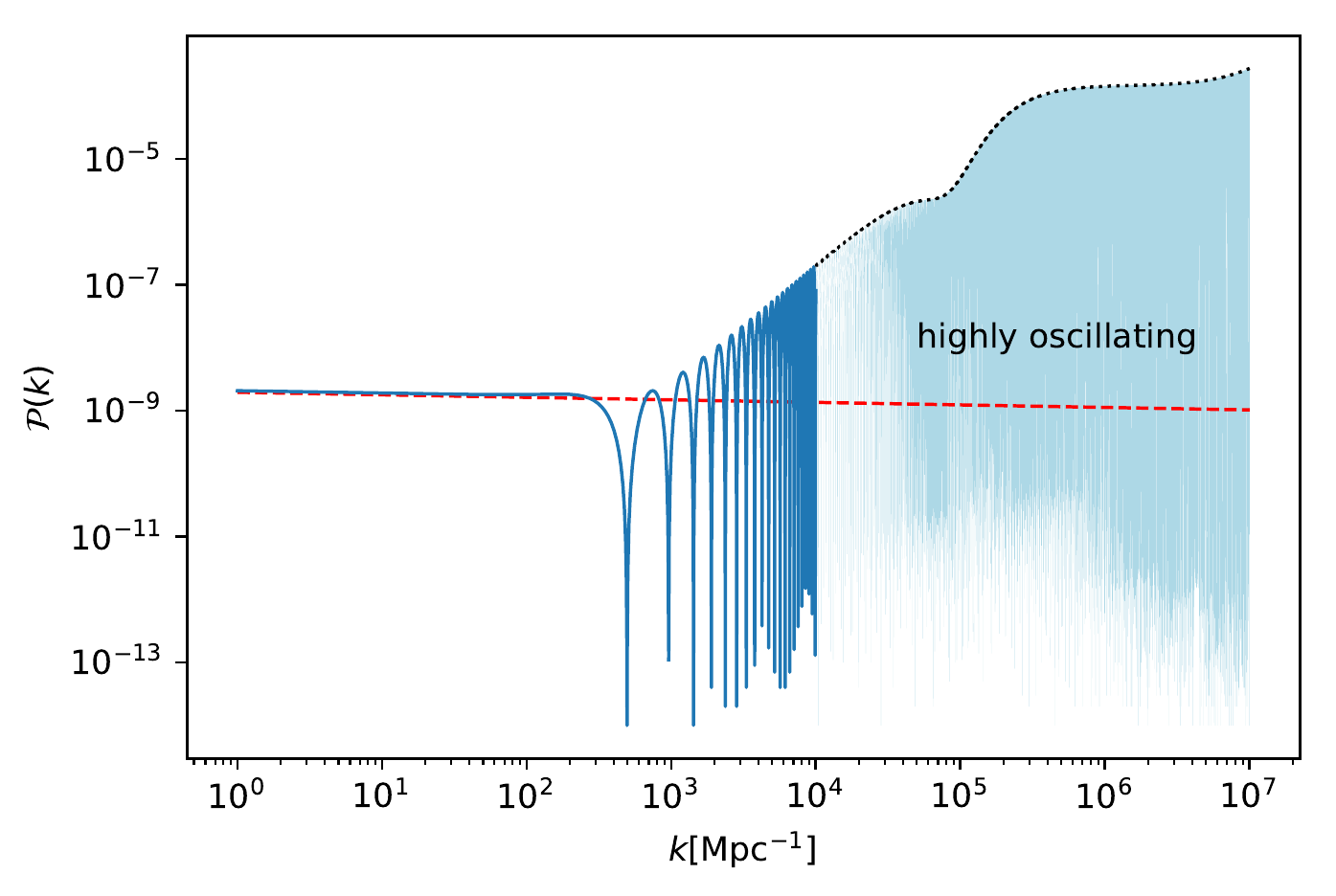}
\caption{Power Spectrum $\fn{\mathcal{P}}{k}$ for $k_{\rm b} = 150 \invMpc$, $k_{\rm a} = 10^5 \invMpc$ and $k_{\rm c} = 10^7 \invMpc$. The red-dotted line represents the power spectrum of $\Lambda$CDM model, i.e. $\mathcal{P}(k) = A_* (k/k_*)^{n_*-1}$.}
\label{ps}
\end{center}
\end{figure}

\fref{transfer} and \fref{ps} are illustrations of the transfer function in \eq{transferfn} and the power spectrum in \eq{Ppsi}. We take $H_{\rm a} \simeq  10^{14} \GeV$ to satisfy the constraint of the energy scale of primordial inflation \citep{Ade:2015lrj}. If $k_{\rm a} \simeq 10^5 \invMpc$ and $H_{\rm b} \simeq 10^6 \GeV$ are taken, we get $N_{\rm ab} \simeq 13$ and $k_{\rm b} \simeq 150 \invMpc$ through \eqs{kakb}{aaab}. If we put $H_{\rm c} \simeq 10^{2} \GeV$, we have $k_{\rm c} = k_{\rm b} e^{N_{\rm bc}} H_{\rm c}/H_{\rm b} \simeq 10^7 \invMpc$ by taking $N_{\rm bc} \simeq 20$. Note that the e-folds during the first inflation are large enough to make the total e-folds through the inflationary phases more than 60. The dips and peaks are related with zeros of Bessel functions. For example, the first dip is found at the first zero of $\fn{J_{\nu'}}{x}$ in \eq{Dpsi}, i.e. $k \simeq 3.14 k_{\rm b} $.

\section{CMB Spectral Distortions}\label{cmb_distortion}

Now we are ready to calculate the CMB spectral distortions generated from \eq{Ppsi} by the full thermalization Green's function \cite{Chluba:2013vsa, Chluba:2015hma}\footnote{We use the numerical code of Green function method which has been developed by Jens Chluba.}. The dissipation of acoustic waves with adiabatic initial conditions heats up the CMB photons generating the spectral distortions. One can calculate this heating rate $\d (Q/\rho_\gamma)/ \d z$  over the redshift $z$, and then the spectral distortion at a given frequency $\Delta I_\nu$ is estimated from the heating rate by
\begin{equation}
\Delta I_\nu \simeq \int \fn{G_{\rm th}}{\nu, z'}  \frac{\d(Q/\rho_\gamma)}{\d z'} \d z'
\end{equation}
where $ \fn{G_{\rm th}}{\nu, z'} $ includes the relevant thermalisation physics, which is independent of the energy release scenario. The spectral distortions is expressed in terms of the temperature shift $\Delta T$, $y$ and $\mu$ contributions 
\begin{equation}\label{distortions}
\Delta I_\nu \approx \frac{\Delta T}{T} \fn{G}{\nu} + y \fn{Y_\mathrm{SZ}}{\nu} + \mu \fn{M_\mathrm{SZ}}{\nu}.
\end{equation}
where $\fn{G}{\nu} = T \frac{\partial \fn{B}{\nu}}{\partial T}$ with $\fn{B}{\nu} = \frac{2h \nu^3}{c^2}\left(e^{x} - 1\right)^{-1}$ for $x = \frac{h \nu}{k_B T} $, $\fn{Y_{\rm SZ}}{\nu} \simeq T \frac{\partial \fn{B}{\nu}}{\partial T} \left( x \fn{\coth}{x}-4  \right)$ and $\fn{M_\mathrm{SZ}}{\nu} \simeq T \frac{\partial \fn{B}{\nu}}{\partial T} \left( 0.4561 - \frac{1}{x}  \right)$.

We should impose $k_{\rm b} \gtrsim 10 \invMpc$ from large scale observations. 
The power spectrum is enhanced around $k = k_{\rm b}$, and it could conflict with CMB observations unless $k_{\rm b}$ should be larger than $10 \invMpc$ \cite{Ade:2015xua}. Primordial balck holes constrain the power spectrum as  $\mathcal{P} \lesssim 10^{-1}$ for all scales  \cite{Josan:2009qn, Emami:2017fiy}. We observe that the enhancement of $\fn{\mathcal{P}}{k}$ tends to grow as $k$ gets large. For $k_{\rm b} \sim 10 \invMpc$, $\fn{\mathcal{P}}{k_{\rm c}} \lesssim 10^{-1}$ can be attained by restricting  $k_{\rm c} \lesssim 10^{8}$.

\begin{figure}
\begin{center}
\includegraphics[width=0.85\textwidth]{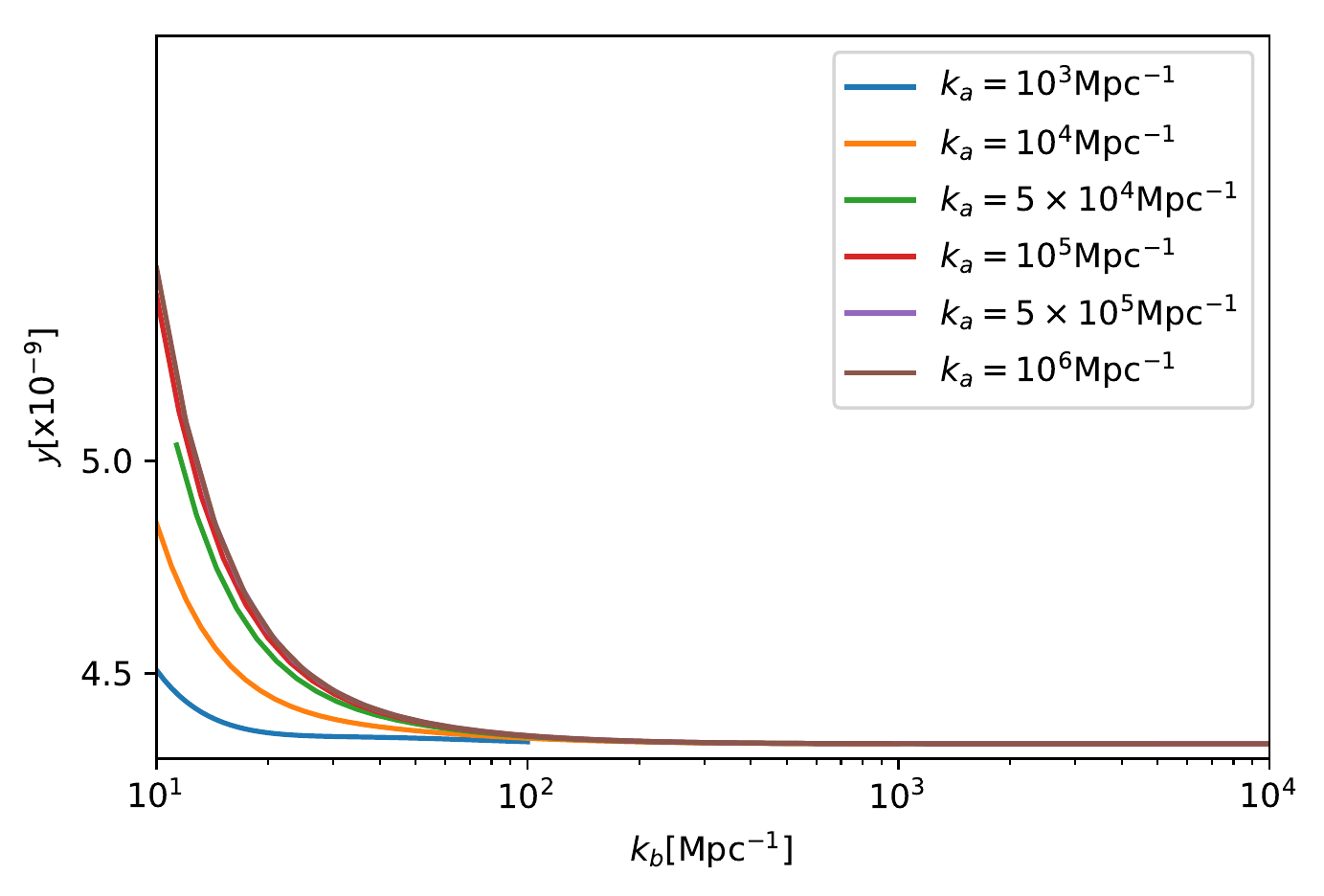}
\caption{$y$-distortion for various values of $k_{\rm a}$ and $k_{\rm c} = 10^7 \invMpc$. We take $10 \invMpc \lesssim k_{\rm b} \lesssim 0.1 k_{\rm a}$.}
\label{y}
\end{center}
\end{figure}

The abundance of ultracompact minihalos (UCMHs) for neutralino dark matter gives  $\mathcal{P} \lesssim 10^{-6}$ at $5 \invMpc \lesssim k \lesssim 10^7 \invMpc$ \cite{Bringmann:2011ut}. (1) For $k_{\rm a} \lesssim 10^7 \invMpc$, we can satisfy $\fn{\mathcal{P}}{k_{\rm a}} < 10^{-6}$ only if $k_{\rm b}$ is not so smaller than $k_{\rm a}$. In this case, the power spectrum is not enhanced within the spectral distortion window  $10 \invMpc \lesssim k \lesssim  10^4 \invMpc$. Thus, the spectral distortions of this case is not distinguished from those of $\Lambda$CDM case. (2) For $k_{\rm a} > 10^7 \invMpc $, the power spectrum may not be constrained by UCMHs, but $H_{\rm a} / H_{\rm b}$ and $N_{\rm ab}$  in \eq{aaab} would be unrealistically huge to enhance the spectral distortions. For example, we need $H_{\rm a} / H_{\rm b} \sim 10^{18}$ and $N_{\rm ab} \sim 30$ for $k_{\rm a} = 10^7 \invMpc$ and $k_{\rm b} = 10 \invMpc$. Hence, UCMHs do not allow the proper ranges for the characteristic scales to distinguish the spectral distortions of our scenario from those of $\Lambda$CDM \cite{Bringmann:2011ut}. However, this constraint is valid when the dark matter is a weakly interacting massive particle (WIMP). If  the dark matter is made up of axions whose masses are within the typical range $10^{-6} \mathrm{eV} \lesssim  m_a \lesssim 10^{-3} \mathrm{eV}$ obtained by astrophysical and cosmological data (e.g. \cite{Bradley:2003kg, Peccei:2006as, Kim:2008hd, Graham:2015ouw, Patrignani:2016xqp}), they cannot annihilate to produce active gamma-rays as WIMPs do. Note that the interactions of these low mass axions are substantially suppressed by the Peccei-Quinn symmetry breaking scale $10^9\GeV \lesssim f_a \lesssim 10^{12} \GeV$, while WIMP interactions are suppressed by the weak interaction scale $m_W \sim m_Z \sim 10^2 \GeV$ \cite{Kim:2008hd}. 
Hence, UCMHs for axion dark matter cannot constrain the power spectrum, and we do not consider the constraints from \cite{Bringmann:2011ut} in our estimation of the spectral distortions. We summarize the ranges of the characteristic scales as
\begin{equation}
10\invMpc \ll k_{\rm b} < k_{\rm a} <  k_{\rm c} \lesssim 10^8 \invMpc ~,
\end{equation}
and could expect that CMB spectral distortions is enhanced within these ranges.

\begin{figure}
\begin{center}
\includegraphics[width=0.85\textwidth]{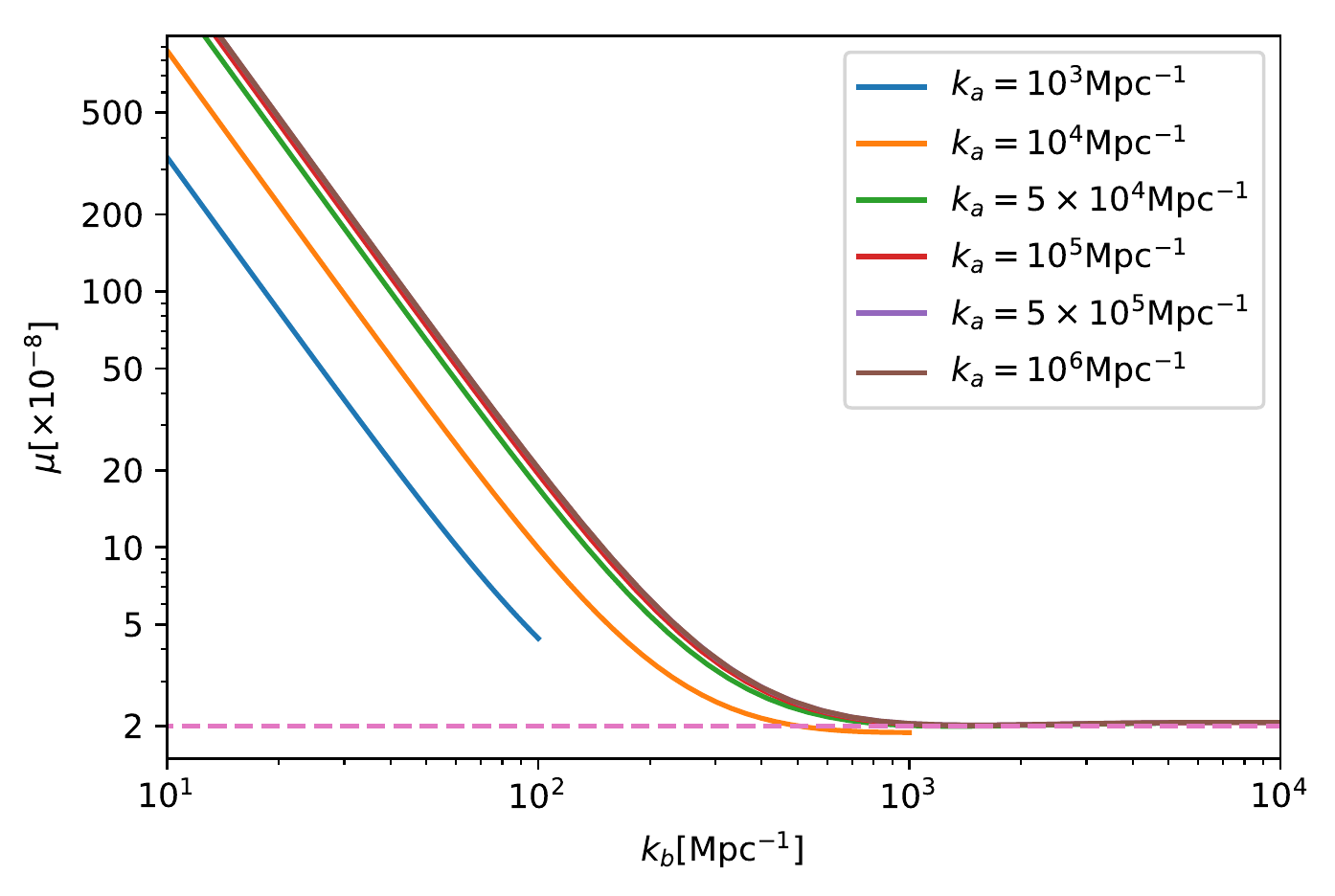}
\caption{$\mu$-distortions for various values of $k_{\rm a}$ and $k_{\rm c} = 10^7 \invMpc$. We take $10 \invMpc \lesssim k_{\rm b} \lesssim 0.1 k_{\rm a}$. The red-dotted line represents the $\mu$-value of $\Lambda$CDM model. }
\label{mu}
\end{center}
\end{figure}

$y$- and $\mu$-distortions for a few examples are shown in \fref{y} and \fref{mu}. For $\mu$-distortion, we include a small correction $\Delta \mu \simeq -0.334 \times 10^{-8}$ due to the energy extraction from photons to baryons as CMB  photons heat up the non-relativistic plasma of baryons by Compton scattering \cite{Chluba:2011hw}. Both $\mu$- and $y$-distortions are safely below than the limit of COBE/FIRAS for $10 \invMpc \lesssim k_{\rm b} $ \cite{Mather:1993ij}. 

We see that $y$-value hardly changes over the range of $k_{\rm b}$ in \fref{y}, while $\mu$-distortion clearly depends on the value of $k_{\rm b}$ in \fref{mu}. $y$-distortion is sensitive to the power spectrum at $k \lesssim 50 \invMpc$  \cite{Chluba:2012we}, but the power spectrum is enhanced at $k > k_{\rm b} > 10 \invMpc$. Then, $y$-distortion is generated by the dissipation of the modes within a very small range $k_{\rm b} \lesssim k \lesssim 50 \invMpc$. As $k_{\rm a}$ gets large, the power spectrum could be enhanced in the range, and $y$-distortion gets slightly large in \fref{y}. However, $\mu$-distortion is sensitive to the power spectrum at $50 \invMpc \lesssim k \lesssim 10^4 \invMpc$  \cite{Chluba:2012we}. When $k_{\rm a}$ is large, the power spectrum could be substantially enhanced in the range, and $\mu$-distortion can be $10^2$ times larger than  the value of $\Lambda$CDM $\mu_{\rm \Lambda CDM}  \simeq 2 \times 10^{-8}$ in \fref{mu}. Therefore, the observation of $\mu$-value is a key to test the enhancement of power spectrum on small scales in our multiple inflation scenario.

\section{Discussion}\label{dis}

In this paper, we consider multiple inflation scenario with having an intermediate matter domination between inflations in \fref{scales}, and find that the power spectrum is enhanced at $k > k_{\rm b}$ where $k_{\rm b}$ is defined as the comoving horizon scale at the beginning of the second inflation by \eq{kb}. If we require $k_{\rm b} > 10 \invMpc$, the enhancement of power spectrum at $k > k_{\rm b}$ has no trouble with large scale observations such as CMB. For $k_{\rm b} > 10^3 \invMpc$, the spectral distortions of the multiple inflation scenario are expected to be similar to that of the standard $\Lambda$CDM model. For $k_{\rm b} \lesssim 10^3 \invMpc$, however, $\mu$-distortion value would be allowed within the COBE/FIRAS, but could be $10^2$ times larger than the expectation of $\Lambda$CDM model which future distortion experiments may be able to test.  

We do not cover various topics on the astrophysical and cosmological applications of the multiple inflation scenario in this paper. First, there could be constraints on the energy scale of the second inflation from particle physics and cosmology, especially in the context of supersymmetric model \cite{German:2001tz,Ross:2010fg,Kaneta:2017lnj}. Two inflations at different energy scales may produce a special profile of gravitational waves background \cite{Zelnikov:1991nv,Polarski:1995zn,Choi:2014aca}. Second, the enhanced power spectrum may impact on the halo abundance and galaxy substructure as in \cite{Hong:2017knn}. This is entangled with the issues such as primordial black holes, minihalos and 21cm observations, which may also give feedbacks to $\mu$-distortions \cite{Tashiro:2008sf,Pani:2013hpa,Kohri:2014lza,Gong:2017sie,Nakama:2017xvq}. If we consider those issues altogether, then multiple inflation scenario would be more constrained. We leave these aspects as future work.

\section*{Acknowledgements}
The authors thank to Ewan Stewart, Kihyun Cho, Sungwook Hong, Hassan Firouzjahi and  Jens Chluba for helpful discussions. HZ thanks Subodh Patil for useful discussions stimulating this work.  We use Jens Chluba's Green function code to estimate CMB spectral distortions. This work is supported by the DGIST  UGRP grant.

\bibliographystyle{apsrev4-1}
\bibliography{Small_Scale_Perturbation}

\end{document}